\documentclass[12pt]{article}
\usepackage{epsfig}
\usepackage{latexsym}
\usepackage{cite}

\usepackage{pslatex}
\usepackage[latin1]{inputenc}
\usepackage[T1]{fontenc}

\usepackage{color,colordvi}
\def\colour4colour#1{\Blue{#1}}

\newcommand{\beq}{\begin{equation}}
\newcommand{\eeq}{\end{equation}}
\newcommand{\bea}{\begin{eqnarray}}
\newcommand{\eea}{\end{eqnarray}}
\newcommand{\nn}{\nonumber}
\newcommand{\MSb}{$\overline{\mbox{MS}}$}
\newcommand{\als}{\alpha_{\rm s}}
\newcommand{\as}{a_{\rm s}}
\newcommand{\ra}{\rightarrow}
\newcommand{\hspn}{{\hspace{-3mm}}}
\newcommand{\ep}{\epsilon}

\begin{document}
\setlength{\parskip}{0.2cm}
\setlength{\baselineskip}{0.54cm}

\def\plus{{\!+\!}}
\def\minus{{\!-\!}}
\def\z#1{{\zeta_{#1}}}
\def\zs{{\zeta_{2}^{\,2}}}
\def\ca{{C^{}_A}}
\def\cf{{C^{}_F}}
\def\cfs{{C^{\, 2}_F}}
\def\cft{{C^{\, 3}_F}}
\def\nf{{n^{}_{\! f}}}
\def\nfs{{n^{\,2}_{\! f}}}

\def\pqq(#1){p_{\rm{qq}}(#1)}
\def\H(#1){{\rm{H}}_{#1}}
\def\Hh(#1,#2){{\rm{H}}_{#1,#2}}
\def\Hhh(#1,#2,#3){{\rm{H}}_{#1,#2,#3}}
\def\Hhhh(#1,#2,#3,#4){{\rm{H}}_{#1,#2,#3,#4}}

\begin{titlepage}
\noindent
DESY 06-036 \hfill {\tt hep-ph/0604053}\\
DCPT/06/40, $\,\,$IPPP/06/20\\[1mm]
April 2006 \\
\vspace{1.8cm}
\begin{center}
\LARGE
{\bf Next-to-Next-to-Leading Order Evolution } \\[2mm]
{\bf of Non-Singlet Fragmentation Functions} \\
\vspace{2.2cm}
\large
A. Mitov$^{\, a}$, S. Moch$^{\, a}$ and A. Vogt$^{\, b}$\\
\vspace{1.2cm}
\normalsize
{\it $^a$Deutsches Elektronensynchrotron DESY \\
\vspace{0.1cm}
Platanenallee 6, D--15735 Zeuthen, Germany}\\
\vspace{0.5cm}
{\it $^b$IPPP, Department of Physics, University of Durham\\
\vspace{0.1cm}
South Road, Durham DH1 3LE, United Kingdom}\\
\vfill
\large
{\bf Abstract}
\vspace{-0.2cm}
\end{center}
We have investigated the next-to-next-to-leading order (NNLO) corrections to 
inclusive hadron production in $e^+e^-$ annihilation and the related parton 
fragmentation distributions, the `time-like' counterparts of the `space-like' 
deep-inelastic structure functions and parton densities.
We have re-derived the corresponding second-order coefficient functions in 
massless perturbative QCD, which so far had been calculated only by one group. 
Moreover we present, for the first time, the third-order splitting functions 
governing the NNLO evolution of flavour non-singlet fragmentation distributions.
These results have been obtained by two independent methods relating time-like 
quantities to calculations performed in deep-inelastic scattering. 
We briefly illustrate the numerical size of the NNLO corrections, and make a 
prediction for the difference of the yet unknown time-like and space-like 
splitting functions at the fourth order in the strong coupling constant.
\vspace{1.0cm}
\end{titlepage}
In this letter we address the evolution of the parton fragmentation 
distributions $D^{\,h}$ and the corresponding fragmentation functions 
$F_a^{\:\! h}\,$ in $e^+e^-$ annihilation, $\, e^+e^- \,\ra\, \gamma,\,Z 
\,\ra\, h + X$ where
\beq
\label{eq:d2sigma}
  {1 \over \sigma_{\:\!\rm tot}}\, {d^{\:\!2} \sigma \over dx\: d\!\cos \theta} 
  \; = \;  {3 \over 8} (1 + \cos^2 \theta) \; F_T^{\:\! h} 
         + {3 \over 4} \: \sin^2 \theta \; F_L^{\:\! h} 
         + {3 \over 4} \: \cos \theta \; F_A^{\:\! h}
  \:\: .
\eeq
Here $\theta$ represents the angle (in the center-of-mass frame) between the
incoming electron beam and the hadron $h$ observed with four-momentum $p$,
and the scaling variable reads $x = 2pq/Q^2$ where $q$ with $q^2\equiv Q^2 > 0$
is the momentum of the virtual gauge boson. The transverse ($T$), longitudinal
($L$) and asymmetric ($A$) fragmentation functions in Eq.~(\ref{eq:d2sigma})
have been measured especially at LEP, see Ref.~\cite{Biebel:2001ka} for a 
general overview. 
Disregarding corrections suppressed by inverse powers of $Q^2$, these 
observables are related to the universal fragmentation distributions $D^{\,h}$ 
by
\beq
\label{eq:Fah}
  F_a^{\:\! h}(x,Q^2) \; = \; \sum_{\rm f = q,\,\bar{q},\,g} \; 
  \int_x^1 {dz \over z} \: c^{}_{a,{\rm f}} \left( z,\als (Q^2) \right) 
  \:  D_{\rm f}^{\,h} \Big( {x \over z},Q^2 \Big)
  \:\: .
\eeq
The coefficient functions $c_{a,{\rm f}}$ in Eq.~(\ref{eq:Fah}) have been 
calculated by Rijken and van Neerven in Refs.\ 
\cite{Rijken:1996vr,Rijken:1996ns,Rijken:1996np} up to the
next-to-next-to-leading order (NNLO) for Eq.~(\ref{eq:d2sigma}), i.e., the 
second order in the strong coupling $\as \equiv \als(Q^2)/(4\pi)$. 
Below we will present the results of a re-calculation of these 
functions by two approaches differing from that employed  
in Refs.~\cite{Rijken:1996vr,Rijken:1996ns,Rijken:1996np}. 

Besides the second-order coefficient functions, a complete NNLO description
also requires the third-order contributions to the splitting functions 
(so far calculated only up to the second order \cite{Curci:1980uw,%
Furmanski:1980cm,Floratos:1981hs}$\,$) governing the scale dependence 
(evolution) of the parton fragmentation distributions. In a notation covering 
both the (time-like $q$, $\sigma=1$) fragmentation distributions and the 
(space-like~$q$, $Q^2 \equiv - q^2$, $\sigma=-1$) parton distributions, the 
flavour non-singlet evolution equations read
\beq
\label{eq:nsevol}
  {d \over d \ln Q^2} \: f_{\sigma}^{\,\rm ns} (x,Q^2) \; = \;
  \int_x^1 {dz \over z} \: P^{\,\rm ns}_{\sigma} \left( z,\als (Q^2) \right)
  \:  f_{\sigma}^{\,\rm ns} \Big( {x \over z},Q^2 \Big) 
\eeq
with
\beq
\label{eq:Pns}
  P^{\,\rm ns}_{\sigma} \left( x,\als (Q^2) \right) \; = \;
  \as \, P^{(0)\,\rm ns}(x) \: +\: \as^{\:\!2} \, P^{(1)\,\rm ns}_{\sigma}(x) 
  \: +\: \as^{\:\!3} P^{(2)\,\rm ns}_{\sigma}(x) \: +\: \ldots \:\: .
\eeq   
The superscript `ns' in Eqs.~(\ref{eq:nsevol}) and (\ref{eq:Pns}) stands for 
any of the following three types of combinations of (parton or fragmentation) 
quark distributions,
\beq
\label{eq:qns}
  f_{ik}^{\,\pm} \; = \; q_i \pm \bar{q}_i \: - \: (q_k \pm \bar{q}_k)
  \:\: , \quad
  f^{\,\rm v} \; = \; {\textstyle \sum_{r=1}^{\nf} } (q_r - \bar{q}_r) \:\: ,
\eeq
where $\nf$ denotes the number of active (effectively massless) flavours. 
As detailed below, we have\\[-1mm] 
obtained the so far unknown time-like NNLO 
splitting functions $P^{(2)\,\rm ns}_{\sigma=1}(x)$ in Eq.~(\ref{eq:Pns}).

As already indicated in Eq.~(\ref{eq:Pns}), the space-like and time-like 
non-singlet splitting functions are identical at the leading order (LO) \cite
{Gribov:1972ri}, a fact known as the Gribov-Lipatov relation. This relation 
does not hold beyond LO in the usual \MSb\ scheme adopted also in this letter.
However, the space-like and time-like cases are related by an analytic 
continuation in $x$, as shown in detailed diagrammatic analyses 
\cite{Curci:1980uw,Stratmann:1996hn} at order $\als^2$, see also Refs.~\cite
{Broadhurst:1993ru,Blumlein:2000wh}. 
Moreover, another approach relating the non-singlet splitting functions has 
been proposed in Ref.~\cite{Dokshitzer:2005bf}. Hence it should be possible to 
derive time-like quantities from the space-like results computed to order 
$\als^{\,3}$ in Refs.~\cite{Moch:2004pa,Vogt:2004mw,Vermaseren:2005qc}.

We start the analytic continuation from the unrenormalized (and unfactorized)
partonic transverse structure function $F_1^{\,\rm b}$ in deep-inelastic 
scattering, $\gamma^{\,\ast} q \ra X$ (and correspondingly for $F_L$ and $F_3 
\ra F_A$), calculated in dimensional regularization with $D = 4 - 2\ep$ and the 
scale $\mu$ \cite{Moch:2004pa,Vogt:2004mw,Vermaseren:2005qc}, 
\beq
\label{eq:F1b}
  F_1^{\,\rm b} (\as^{\,\rm b}, Q^2) \; = \; 
  \delta(1-x) \:  + \: \sum_{l=1}^{\infty} ( \as^{\,\rm b\:\!} )^l
  \left( Q^2 \over \mu^2 \right)^{-l\ep} F_{1,l}^{\,\rm b} \:\: .
\eeq
The bare and renormalized coupling $\als^{\,\rm b}$ and $\als$ are related by
(recall $\as \equiv \als/(4\pi)\:\!$) 
\beq
\label{eq:asb}
  \as^{\,\rm b} \; = \; \as \: -\: {\beta_0 \over \ep}\, \as^{\:\!2} 
  \: +\: \left( {\beta_0^2 \over \ep^2} - {\beta_1 \over 2\ep} \right)
  \as^{\:\!3}
  \: +\: \ldots
\eeq
with $\,\beta_0 = 11/3\; \ca - 2/3\; \nf\,$ etc. The expansion coefficients in
Eq.~(\ref{eq:F1b}) are then decomposed into form-factor (${\cal F}_l$) and 
real-emission (${\cal R}_{\,l}$, defined analogous to Eq.~(\ref{eq:F1b})$\:\!$) 
contributions \cite{Moch:2005id}
\bea
\label{eq:F1dec}
F_{1,1}^{\,\rm b}
     &\!=\!& 2 {\cal F}_1\,\delta(1-x) + {\cal R}_{\,1} \nn \\[0.5mm]
F_{1,2}^{\,\rm b}
     &\!=\!& 2 {\cal F}_2\, \delta(1-x)
           + \left({\cal F}_1\right)^2 \delta(1-x)
           + 2 {\cal F}_1 {\cal R}_{\,1} + {\cal R}_{\,2} \nn \\[0.5mm]
F_{1,3}^{\,\rm b}
     &\!=\!& 2 {\cal F}_3\, \delta(1-x)
           + 2 {\cal F}_1 {\cal F}_2\, \delta(1-x)
           + (\, 2 {\cal F}_2 + \left({\cal F}_1\right)^2 \,)\, {\cal R}_{\,1} 
           + 2 {\cal F}_1 {\cal R}_{\,2} + {\cal R}_{\,3} 
\:\: .
\eea
  
The analytic continuation of the form factor to the time-like case is known.
The $x$-dependent functions ${\cal R}_l$ are continued from $x$ to $1/x$
\cite{Curci:1980uw,Stratmann:1996hn}, taking into account the (complex) 
continuation of $q^2$ (see Eq.~(4.1) of Ref.~\cite{Moch:2005id}$\,$) and the
additional prefactor $x^{1-2\ep}$ originating from the phase space of the 
detected parton in the time-like case \cite{Rijken:1996ns}. 
Practically this continuation has been performed using routines for harmonic 
polylogarithms \cite{Remiddi:1999ew,Moch:1999eb} implemented in {\sc Form} 
\cite{Vermaseren:2000nd}. The only subtle point in the analytic continuations
is the treatment of logarithmic singularities for $x \ra 1$, cf.~Ref.~\cite
{Stratmann:1996hn}, starting with 
\beq
\label{eq:l1xcnt}
  \ln (1-x) \;\ra\; \ln (1-x) - \ln x + i\,\pi \:\: .
\eeq
 
Finally the bare transverse fragmentation function $F_T^{\,\rm b}$ is 
re-assembled analogous to Eq.~(\ref{eq:F1dec}), keeping the real parts of the 
continued ${\cal R}_{\,l}$ only, and the time-like non-singlet splitting 
functions and coefficient functions can be read off iteratively from the 
non-singlet mass factorization formula
\bea
\label{eq:FTfact}
 F^{}_{T,1} &\!=\!& \!
   - \,\ep^{-1}\, P^{(0)} \: +\: c^{(1)}_{T} \: +\: \ep\, a^{(1)}_{T} 
   \: +\: \ep^2\, b^{(1)}_{T} \: +\: \ep^3\, d^{(1)}_{T} \: +\: \ldots \nn \\
 F^{}_{T,2} &\!=\!& \!
     {1 \over 2\ep^2}\, P^{(0)} ( P^{(0)} + \beta_0 ) 
   - {1 \over 2\ep}\, \Big[ P_{\sigma=1}^{(1)} + 2 P^{(0)} c^{(1)}_{T} \Big]
   + c^{(2)}_{T} - P^{(0)} a^{(1)}_{T} 
   + \ep\, \Big[ a^{(2)}_{T} - P^{(0)} b^{(1)}_{T} \Big] + \ldots \nn \\ 
 F^{}_{T,3} &\!=\!& \!
   - {1 \over 6\ep^3}\, P^{(0)} ( P^{(0)} + \beta_0 ) ( P^{(0)} + 2 \beta_0 )
   + {1 \over 6\ep^2}\, \Big[ P_{\sigma=1}^{(1)} ( 3 P^{(0)} + 2\beta_0 )
     + P^{(0)} ( 3 P^{(0)} c^{(1)}_{T} 
\\ & & \; \mbox{} 
     + 3\beta_0 c^{(1)}_{T} + 2\beta_1 ) \Big]
   - {1 \over 6\ep}\, \Big[ 2 P_{\sigma=1}^{(2)} 
     + 3 P_{\sigma=1}^{(1)} c^{(1)}_{T} + P^{(0)} 
     ( 6 c^{(2)}_{T} - 3 P^{(0)} a^{(1)}_{T} - 3\beta_0 a^{(1)}_{T} ) \Big]
     + \ldots \nn
\eea
where the expansion coefficients $F^{}_{T,l}$ now refer to an expansion in the
renormalized coupling at the scale $Q^2 \equiv q^2$. The products of the 
$x$-dependent (generalized) functions in Eq.~(\ref{eq:FTfact}) are to be read
as Mellin convolutions or, more conveniently, as products in Mellin-$N$ space,
employing routines for harmonic sums and their inverse Mellin transform back
to $x$-space \cite{Vermaseren:1998uu,Moch:1999eb,Vermaseren:2000nd}.

Unlike the diagrammatic treatment of Refs.~\cite{Curci:1980uw,Stratmann:1996hn}
--- which cannot be emulated at order $\als^{\,3}$ using the only available
space-like calculation based on the optical theorem 
\cite{Moch:2004pa,Vogt:2004mw,Vermaseren:2005qc} ---
the above procedure is not entirely rigorous, as Eq.~(\ref{eq:F1dec}) does not
represent a full decomposition according to the number of emitted partons.
This would be required, for instance, in a subtraction formalism for exclusive 
observables at NNLO \cite{Frixione:2004is} and beyond.
In Eq.~(\ref{eq:F1dec}) the real functions ${\cal R}_{\:n\geq 2}$ do not only 
collect, e.g., $n$-gluon tree-level amplitudes, but also combinations of real 
emission and virtual corrections.
Especially, starting at order $\als^{\,3}$, the decomposition 
Eq.~(\ref{eq:F1dec}) includes overlapping divergences from triple unresolved 
configurations when two particles become soft and one is collinear.
Thus one has to be prepared for some problem in the abelian ($C_F^{\,3}$) piece
related to $\pi^2$ contributions originating from phase space integrations
over unresolved regions.

At the second order, however, the above procedure works perfectly at least for 
the terms in the $\ep$-expansion written down in the second line of 
Eq.~(\ref{eq:FTfact}), 
thus including the previously unknown \mbox{$\ep$-coefficients} $a_a^{(2)}(x)$, 
$a = T,\:L,\:A$, which we will not write out here for brevity.
We have verified this by comparing to a direct calculation, to be presented 
elsewhere, of $\, e^+e^- \,\ra\, \gamma,\,Z \,\ra\, h + X$ to this 
accuracy in $\ep$ using the approach of Ref.~\cite{Mitov:2005ps}. 
Especially, we also re-derive the ${\cal O}(\as^{\,2})$ coefficient functions 
$c_a^{(2)}(x)$ which so far were only calculated in Refs.~\cite
{Rijken:1996vr,Rijken:1996ns,Rijken:1996np}. The differences between the 
time-like non-singlet coefficient functions for Eq.~(\ref{eq:d2sigma}) and the 
corresponding (by the structure of the respective hadronic tensors) quantities 
in deep-inelastic scattering read
\bea
\lefteqn{c^{\,(2)}_{T,\,\rm ns}(x) 
  - c^{\,(2)}_{1,\,\rm ns}(x)\;\; = \;\; }
  \nonumber\\&& \mbox{\hspn}
  + \colour4colour{\cfs}  \*  \Big(
            72
          - 8 \* \z2
          + 176/3\: \* \H(0)
          - 48 \* \Hh(0,0)
          + 8 \* \H(2)
          - 72 \* \Hh(1,0)
       + (1+x) \* \Big[
          - 493/6
          + 12 \* \z3
\nn\\&& \mbox{}
          + 10 \* \z2
          - 155/6\: \* \H(0)
          + 16 \* \H(0) \* \z2
          - 6 \* \Hh(0,0)
          + 12 \* \Hhh(0,0,0)
          - 20 \* \H(3)
          - 10 \* \H(2)
          - 4 \* \Hh(2,0)
          - 29 \* \H(1)
\nn\\&& \mbox{}
          + 42 \* \Hh(1,0)
          - 12 \* \Hh(1,1)
          \Big]
       + \pqq(x)  \*  \Big[
          - 84 \* \z3
          - 106/3\: \* \z2
          + 389/6\: \* \H(0)
          - 44 \* \H(0) \* \z2
\nn\\&& \mbox{}
          + 196/3\: \* \Hh(0,0)
          - 24 \* \Hhh(0,0,0)
          + 36 \* \H(3)
          + 62 \* \H(2)
          - 20 \* \Hh(2,0)
          + 48 \* \Hh(2,1)
          - 40 \* \H(1) \* \z2
          + 134/3\: \* \Hh(1,0)
\nn\\&& \mbox{}
          + 32 \* \Hhh(1,0,0)
          + 56 \* \Hh(1,2)
          + 40 \* \Hhh(1,1,0)
          \Big]
       + \delta(1 - x) \* \Big[
            608/3\: \* \z2
          - 24 \* \zs
          \Big]
          \Big)
\nn\\&& \mbox{\hspn}
  + \colour4colour{\cf\, \* (\ca - 2 \* \cf)}  \*  \Big(
         - 76/3\: \* x \* \H(0)
       + (1+x) \* \Big[
          - 215/6
          + 49/6\: \* \H(0)
          - 9 \* \H(1)
          \Big]
       + \pqq(x) \* \Big[
            12 \* \z3
\nn\\&& \mbox{}
          - 44/3\: \* \z2
          + 445/6\: \* \H(0)
          - 20 \* \H(0) \* \z2
          + 44/3\: \* \Hh(0,0)
          + 24 \* \Hhh(0,0,0)
          + 4 \* \H(3)
          + 22 \* \H(2)
          - 4 \* \Hh(2,0)
\nn\\&& \mbox{}
          - 16 \* \H(1) \* \z2
          + 22/3\: \* \Hh(1,0)
          + 8 \* \Hh(1,2)
          - 8 \* \Hhh(1,1,0)
          \Big]
       + \pqq( - x)  \*  \Big[
            32 \* \Hhh(-1,0,0)
          + 8 \* \H(0) \* \z2
          + 16 \* \Hh(-2,0)
\nn\\[-0.5mm]&& \mbox{}
          - 24 \* \Hhh(0,0,0)
          \Big]
       + \delta(1 - x)  \*  \Big[
            466/3\: \* \z2
          - 24 \* \zs
          \Big]
          \Big)
\nn\\&& \mbox{\hspn}
  + \colour4colour{\cf \* \nf}  \*  \Big(
          4/3\: \* x \* \H(0)
       + (1+x) \* \Big[
            19/3
          + 1/3\: \* \H(0)
          + 2 \* \H(1)
          \Big]
       + \pqq(x) \* \Big[
            8/3\: \* \z2
          - 35/3\: \* \H(0)
\nn\\[-1mm]&& \mbox{}
          - 8/3\: \* \Hh(0,0)
          - 4 \* \H(2)
          - 4/3\: \* \Hh(1,0)
          \Big]
       + \delta(1 - x) \* \Big[
          - 76/3\: \* \z2
          \Big]
          \Big)
\: \: ,\label{eq:dC1T}
\\[1mm]
\lefteqn{c^{\,(2)}_{L,\,\rm ns}(x)|_{e^+e^-} 
  \: - \: c^{\,(2)}_{L,\,\rm ns}(x)|_{ep} \;\; = \;\; } 
  \nonumber\\&& \mbox{\hspn}
  + \colour4colour{\cfs}  \*  \Big(
            583/9
          - 860/9\: \* x
          - (14/3 + 152/3\: \* x) \* \H(0)
          - (12 + 16 \* x) \* \Hh(0,0)
          + (74/3 - 112/3\: \* x) \* \H(1)
\nn\\[-1mm]&& \mbox{}
          - 16 \* (1+x) \* \Hh(1,0)
          + 8 \* (1-2 \* x) \* \Hh(1,1)
          + 4 \* (1+6 \* x) \* \Big[
            \z2
          - \H(2)
          \Big]
          \Big)
\nn\\&& \mbox{\hspn}
  + \colour4colour{\cf\, \* (\ca - 2 \* \cf)}  \*  \Big(
            2317/45
          + 8/5\: \* x^{-1}
          - 3752/45\: \* x
          + 32/5\: \* x^2
          + 32 \* x \* \z3
          + (
            40
          + 8/5 \* x^{-2}) \* \Hh(-1,0)
\nn\\[-1mm]&& \mbox{}
          - 1/5 \* (
            98/3
          + 8 \* x^{-1}
          + 728/3\: \* x
          - 32 \* x^2) \* \H(0)
          + 16 \* \Hhh(0,-1,0)
          + 32 \* ( x
          - 1/5\: \* x^3) \* \Big[
            \z2
          + \Hh(-1,0)
          - \Hh(0,0)
          \Big]
\nn\\[-1mm]&& \mbox{}
          - 8 \* (1+2 \* x) \* \Big[
            \H(-1) \* \z2
          + 2 \* \Hhh(-1,-1,0)
          - \Hhh(-1,0,0)
          \Big]
          + (1-2 \* x) \* \Big[
            46/3\: \* \H(1)
          - 8 \* \H(1) \* \z2
          + 8 \* \Hhh(1,0,0)
          \Big]
          \Big)
\nn\\&& \mbox{\hspn}
  + \colour4colour{\cf \* \nf}  \*  \Big(
          - 74/9 
          + 112/9\: \* x
          + 4/3\: \* (1+4 \* x) \* \H(0)
          - 4/3\: \* (1-2 \* x) \* \H(1)
         \Big)
\: \: ,\label{eq:dCLT}
\\[1mm]
\lefteqn{c^{\,(2)}_{A,\,\rm ns}(x) 
  - c^{\,(2)}_{3,\,\rm ns}(x) 
  \;\; = \;\; c^{\,(2)}_{T,\,\rm ns}(x) 
            - c^{\,(2)}_{1,\,\rm ns}(x) }
  \nonumber\\&& \mbox{\hspn}
  + \colour4colour{\cfs}  \*  \Big(
          (1-x) \* \Big[
          - 16 \* \z2
          + 46 \* \H(0)
          + 20 \* \Hh(0,0)
          + 16 \* \H(2)
          + 24 \* \Hh(1,0)
          \Big]
          \Big)
  \: + \: \colour4colour{\cf \* \nf}  \*  \Big(
          - 4 \* (1-x) \* \H(0)
          \Big)
\nn\\[-1mm]&& \mbox{\hspn}
  + \colour4colour{\cf\, \* (\ca - 2 \* \cf)}  \*  \Big(
         \pqq( - x) \* \Big[
          - 64 \* \Hhh(-1,0,0)
          - 16 \* \H(0) \* \z2
          - 32 \* \Hh(-2,0)
          + 48 \* \Hhh(0,0,0)
          \Big]
\nn\\[-1mm]&& \mbox{}
          + 54 \* (1-x) \* \H(0)
          + 32 \* (1+x) \* \Hh(0,0)
          \Big)
\: \: .\label{eq:ddC3A}
\eea
Here our notation for the harmonic polylogarithms $H_{m_1,...,m_w}(x)$,
$m_j = 0,\pm 1$ follows Ref.~\cite{Remiddi:1999ew}. Furthermore we have
employed the short-hand notation
\bea
\label{eq:habbr}
  H_{{\footnotesize \underbrace{0,\ldots ,0}_{\scriptstyle m} },\,
  \pm 1,\, {\footnotesize \underbrace{0,\ldots ,0}_{\scriptstyle n} },
  \, \pm 1,\, \ldots}(x) & = & H_{\pm (m+1),\,\pm (n+1),\, \ldots}(x)
  \:\: , \\[-5mm] \nn
\eea
suppressed the argument of the polylogarithms, and used the function
$
  \, p_{\rm{qq}}(x) \, = \, 2\, (1 - x)^{-1} - 1 - x \, .
$
The divergences for $x \ra 1$ in Eq.~(\ref{eq:dC1T}) are to be read as
plus-distributions. 

Eqs.~(\ref{eq:dC1T}) - (\ref{eq:ddC3A}) agree with the results in Refs.~\cite
{Rijken:1996np,Rijken:1996ns} up to a few (presumably typographical) errors in 
those articles. Specifically, for $c_T^{(2)}(z)$ the term 
$ \, -3\, \cfs\, (1+z) \ln^2 z $ in Eq.~(A.6) of 
Ref.~\cite{Rijken:1996ns} has to be replaced by
$ \, -3\, \cfs\, (1+z) \ln^3 z $, and the contribution
$ \, 24\, (\cfs- \ca\cf/2 ) \, \ln z / (5 z^2) $
in Eq.~(A.8) by 
$ \, 24\, (\cfs- \ca\cf/2 ) \, \ln z / (5 z) $.
The argument of $S_{1,2}$ should read $\, -z\,$ instead of $\, 1-z\,$ in the
first term of Eq.~(A.15) for $c_L^{(2)}(z)$.  
Finally, in Eq.~(17) of Ref.~\cite{Rijken:1996np} for
$c_A^{(2)}(z)$ the term
$ \, 24\, \cfs \, \ln z / (5 z^2) $
has to be replaced by
$ \, 24\, \cfs \, \ln z / (5 z) $.
We have also re-calculated the second-order gluon and pure-singlet coefficient 
functions for $c_T^{}$ and $c_L^{}$, finding complete agreement with 
Ref.~\cite{Rijken:1996np}.
 
We now turn to the corresponding NNLO (third-order) splitting functions 
$P_{\sigma=1}^{(2)\,\rm ns}$ in Eq.~(\ref{eq:Pns}). Using the same notation as 
above these functions are given by
\bea
  \lefteqn{ \delta\, P^{\,(2)+}(x) \;\; \equiv \;\;
           P^{\,(2)+}_{\sigma=1}(x) 
         - P^{\,(2)+}_{\sigma=-1}(x)\;\; = \;\; }
  \nonumber\\&& \mbox{\hspn}
  + 16 \, \* \colour4colour{\cft}  \*  \Big(
       \pqq(x) \*  \Big[
            311/24\: \* \H(0)
          + 4/3\: \* \H(0) \* \z2
          - 169/9\: \* \Hh(0,0)
          + 8 \* \Hh(0,0) \* \z2
          - 22 \* \Hhh(0,0,0)
\nn\\[-0.8mm]&& \mbox{}
          - 268/9\: \* \Hh(1,0)
          + 8 \* \Hh(1,0) \* \z2
          - 44/3\: \* \Hhh(1,0,0)
          - 268/9\: \* \H(2)
          + 8 \* \H(2) \* \z2
          - 44/3\: \* \Hh(2,0)
          - 44/3\: \* \H(3) \Big]
\nn\\[-0.8mm]&& \mbox{}
       + (1+x) \* \Big[
          - 4 \* \H(0,0) \* \z2
          + 25/2\: \* \H(0,0,0)
          + \H(2,0)
          + 2 \* \H(3)
          \Big]
       - (1-x) \* \Big[
            325/18\: \* \H(0)
          + 50/3\: \* \H(1,0)
\nn\\[-0.8mm]&& \mbox{}
          + 50/3\: \* \H(2)
          \Big]
       + ( 3 - 5 \* x) \*  \H(0) \* \z2
       - ( 173/18 - 691/18\: \* x ) \* \H(0,0) \Big)
\nn\\&& \mbox{\hspn}
  + 16 \, \* \colour4colour{\cfs\, \* (\ca - 2 \* \cf)}  \*  \Big(
       \pqq(x) \*  \Big[
            151/24\: \* \H(0) 
          + \H(0) \* \z3
          + 13/6\: \* \H(0) \* \z2
          - 169/18\: \* \Hh(0,0)
          + 8 \* \Hh(0,0) \* \z2
\nn\\&& \mbox{}
          - 13/2\: \* \Hhh(0,0,0)
          - 8 \* \Hhhh(0,0,0,0)
          - 134/9\: \* \Hh(1,0)
          + 4 \* \Hh(1,0) \* \z2
          - 22/3\: \* \Hhh(1,0,0)
          - 6 \* \Hhhh(1,0,0,0)
          - 134/9\: \* \H(2)
\nn\\&& \mbox{}
          + 4 \* \H(2) \* \z2
          - 22/3\: \* \Hh(2,0)
          - 2 \* \Hhh(2,0,0)
          - 22/3\: \* \H(3)
          - 2 \* \Hh(3,0)
          - 6 \* \H(4) \Big]
       + \pqq(-x) \*  \Big[
          - 8 \* \Hh(-3,0)
\nn\\&& \mbox{}
          + 8 \* \H(-2) \* \z2
          + 8 \* \Hhh(-2,-1,0)
          + 3 \* \Hh(-2,0)
          - 14 \* \Hhh(-2,0,0)
          - 4 \* \Hh(-2,2)
          + 8 \* \Hhh(-1,-2,0)
          + 16 \* \Hhhh(-1,-1,0,0)
\nn\\[1mm]&& \mbox{}
          + 8 \* \Hh(-1,0) \* \z2
          + 6 \* \Hhh(-1,0,0)
          - 18 \* \Hhhh(-1,0,0,0)
          - 4 \* \Hhh(-1,2,0)
          - 8 \* \Hh(-1,3)
          - 7 \* \H(0) \* \z3
          + 3/2\: \* \H(0) \* \z2
\nn\\&& \mbox{}
          - 8 \* \Hh(0,0) \* \z2
          - 9/2\: \* \Hhh(0,0,0)
          + 8 \* \Hhhh(0,0,0,0)
          + 2 \* \Hh(3,0)
          + 6 \* \H(4) \Big]
       - (1+x) \* \Big[
            4 \* \H(-2,0)  
          + 8 \* \H(-1,0,0) \Big]
\nn\\[-1mm]&& \mbox{}
       + (1-x) \* \Big[
            4 \* \H(-3,0)
          + 4 \* \H(-2,0,0)
          - 88/9\: \* \H(0)
          + 3 \* \H(0) \* \z3
          - 28/3\: \* \H(1,0)
          - 28/3\: \* \H(2) \Big]
       - 4 \* x \* \H(0) \* \z2 
\nn\\[-1mm]&& \mbox{}
       - ( 50/9 - 184/9\: \* x ) \* \H(0,0)
       - 4 \* x \* \H(0,0) \* \z2 
       + ( 11/2 + 35/2\: \* x ) \* \H(0,0,0)
       + 8 \* x \* \H(0,0,0,0) \Big)
\nn\\&& \mbox{\hspn}
  + 16 \, \* \colour4colour{\cfs\, \* \nf}  \*  \Big(
       \pqq(x) \* \Big[
          - 11/12\: \* \H(0)
          - 2/3 \* \H(0) \* \z2
          + 11/9\: \* \H(0,0)
          + 2 \* \H(0,0,0)
          + 20/9\: \* \H(1,0)
\nn\\[-1mm]&& \mbox{}
          + 4/3\: \* \H(1,0,0)
          + 20/9\: \* \H(2)
          + 4/3\: \* \H(2,0)
          + 4/3\: \* \H(3) \Big]
       - (1+x)\: \*
            \Hhh(0,0,0)
       + (1-x) \* \Big[
            13/9\: \* \H(0)
\nn\\[-1mm]&& \mbox{}
          + 4/3\: \* \H(1,0)
          + 4/3\: \* \H(2) \Big]
       + ( 8/9 - 28/9\: \* x ) \* \H(0,0) \Big)
\:\: ,
\label{eq:dPqq2p}
\\[-8mm] \nn
\eea
and
\bea
\nn \\[-7mm] 
  \lefteqn{ \delta\, P^{\,(2)\,\xi}(x) 
          - \delta\, P^{\,(2)+}(x) \;\; \equiv \;\; }
  \nonumber\\&& \mbox{\hspn}
  + 16 \, \* \colour4colour{\cfs\, \* (\ca - 2 \* \cf)}  \*  \Big(
       \pqq(-x) \*  \Big[
            16 \* \Hh(-3,0)
          - 16 \* \H(-2) \* \z2
          - 16 \* \Hhh(-2,-1,0)
          - 6 \* \Hh(-2,0)
          + 28 \* \Hhh(-2,0,0)
\nn\\&& \mbox{}
          + 8 \* \Hh(-2,2)
          - 16 \* \Hhh(-1,-2,0)
          - 32 \* \Hhhh(-1,-1,0,0)
          - 16 \* \Hh(-1,0) \* \z2
          - 12 \* \Hhh(-1,0,0)
          + 36 \* \Hhhh(-1,0,0,0)
          + 8 \* \Hhh(-1,2,0)
\nn\\&& \mbox{}
          + 16 \* \Hh(-1,3)
          + 14 \* \H(0) \* \z3
          - 3 \* \H(0) \* \z2
          + 16 \* \Hh(0,0) \* \z2
          + 9 \* \Hhh(0,0,0)
          - 16 \* \Hhhh(0,0,0,0)
          - 4 \* \Hh(3,0)
          - 12 \* \H(4) \Big]
\nn\\[-1mm]&& \mbox{}
       + (1+x) \* \Big[
            8 \* \Hh(-2,0)
          + 16 \* \Hhh(-1,0,0)
          + 8 \* \H(0) \* \z2
          - 4 \* \Hh(2,0)
          - 8 \* \H(3)
          \Big]
       - (1-x) \* \Big[
            8 \* \Hh(-3,0)
          + 8 \* \Hhh(-2,0,0)
          + 10 \* \H(0)
\nn\\[-1mm]&& \mbox{}
          + 6 \* \H(0) \* \z3
          + 4 \* \Hh(0,0) \* \z2
          - 8 \* \Hhhh(0,0,0,0)
          + 8 \* \Hh(1,0)
          + 8 \* \H(2)
          \Big] 
       - ( 10 - 6 \* x ) \* \Hh(0,0)
       - ( 12 + 24 \* x ) \* \Hhh(0,0,0)  \Big)
\label{eq:ddPqq2}
\eea
for both $\xi = -$ and $\xi = \rm v$.
Eq.~(15) is, in fact, not quite the result of the analytic continuation as 
described above, which returns a different coefficient for the term
$\,\cft\,\pqq(x)\,\Hh(0,0)\,\z2\,$ in the first line. We have 
corrected this term by imposing the correct (vanishing) first moment of
$P_{\sigma=1}^{(2)-}$. As it is conceivable that the present form of the
analytic continuation leads to other problems not affecting the first moment
--- recall the discussion in the paragraph below Eq.~(\ref{eq:FTfact}) ---
we obviously need a second, independent confirmation of our new results 
(\ref{eq:dPqq2p}) and (\ref{eq:ddPqq2}).

For this purpose we adopt the approach of Dokshitzer, Marchesini and Salam
\cite{Dokshitzer:2005bf} (see also Appendix B1 of Ref.~\cite{Dokshitzer:1995ev}%
$\,$), where Eq.~(\ref{eq:nsevol}) is rewritten as\footnote
{Ref.~\cite{Dokshitzer:2005bf} also includes a shift in the 
argument of $\als$ in Eq.~(\ref{eq:DMSevol}) which is irrelevant for our purpose
of relating the time-like and space-like results.
Note also that the notation for the $\als$ expansion in Ref.~\cite
{Dokshitzer:2005bf} differs from Eq.~(\ref{eq:Pns}).}
\beq
\label{eq:DMSevol}
  {d \over d \ln Q^2} \: f_{\sigma}^{\,\rm ns} (x,Q^2) \; = \;
  \int_x^1 {dz \over z} \: P^{\,\rm ns}_{\rm univ} \left( z,\als (Q^2) \right)
  \:  f_{\sigma}^{\,\rm ns} \Big( {x \over z},z^{\sigma}Q^2 \Big) \:\: ,
\eeq
and the modified splitting functions $P^{\,\rm ns}_{\!\rm univ}$ are postulated
to be identical for the time-like and space-like cases. Working out the
perturbative expansion of the integrand along the lines of 
Ref.~\cite{Dokshitzer:2005bf} one arrives at a successful `postdiction' for the 
NLO difference $P^{\,(1)\,\rm ns}_{\sigma=1}(x) - P^{\,(1)\,\rm ns}_{\sigma=-1}
(x)$ of Refs.~\cite{Curci:1980uw,Floratos:1981hs} (see Eq.~(4) of Ref.~\cite
{Dokshitzer:2005bf}$\,$) and the new NNLO prediction ($\,\xi = +,\: -,\: \rm v$,
recall Eq.~(\ref{eq:qns}))
\beq
\label{eq:dP2DMS}
  \delta\, P^{\,(2)\,\xi}(x) \; = \;
   2 \left\{ \Big[ \ln x \cdot \widetilde{P}^{\,(1)\,\xi} \Big] 
              \otimes P^{\,(0)}  
          + \Big[ \ln x \cdot P^{\,(0)} \Big]
              \otimes \widetilde{P}^{\,(1)\,\xi} \right\}
\eeq
with $\otimes$ denoting the Mellin convolution (cf.~Eqs.~(\ref{eq:Fah}) and 
(\ref{eq:nsevol})) and
\beq 
  2\,\widetilde{P}^{\,(n)\,\xi}(x) \; = \; 
   P^{\,(n)\,\xi}_{\sigma=1}(x) + P^{\,(n)\,\xi}_{\sigma=-1}(x) 
  \:\: .
\eeq
The evaluation of Eq.~(\ref{eq:dP2DMS}), again performing the convolutions via a
transformation to Mellin-$N$ space, yields exactly Eqs.~(\ref{eq:dPqq2p}) and 
(\ref{eq:ddPqq2}), thus providing both the desired confirmation of these results
and further evidence supporting the ansatz (\ref{eq:DMSevol}).
 
Consequently, it is possible to make even a prediction for the fourth-order 
(N$^3$LO) difference $\delta\, P^{\,(3)\,\xi}$ of the (both unknown) time-like 
and space-like non-singlet splitting functions on this basis. Using the notation
of Eq.~(\ref{eq:dP2DMS}) together with $A^{\otimes 2} \equiv A \otimes A$ etc, 
this prediction reads
\bea
  \delta\, P^{\,(3)\,\xi}(x) & = & 
  2 \left\{ [\, \ln x \cdot \widetilde{P}^{\,(2)\,\xi} \,] 
              \otimes P^{\,(0)}  
          + [\, \ln x \cdot P^{\,(0)} \,]
              \otimes \widetilde{P}^{\,(2)\,\xi} 
          + [\, \ln x \cdot \widetilde{P}^{\,(1)\,\xi} \,]
              \otimes \widetilde{P}^{\,(1)\,\xi} 
    \right\}
 \nn\\ & & \mbox{\hspn}
 - 2\: P^{\,(0)} \otimes [\, \ln x \cdot P^{\,(0)} \,]^{\otimes 3} 
 - 4\: [\, P^{\,(0)} \,]^{\otimes 2} \otimes [\, \ln x \cdot P^{\,(0)} \,]
   \otimes [\, \ln^2 x \cdot P^{\,(0)} \,]
 \nn\\[1mm] & & \mbox{\hspn}
 - 2/3\; [\, P^{\,(0)} \,]^{\otimes 3} \otimes [\, \ln^3 x \cdot P^{\,(0)} \,]
 \:\: .
\label{eq:dP3DMS}
\eea
For brevity we refrain from writing out the resulting explicit expressions
which are, of course, much more lengthy than Eq.~(\ref{eq:dPqq2p}).
We expect that a first check of Eq.~(\ref{eq:dP3DMS}), or rather its first line
(which dominates the large-$x$ behaviour, cf.~Ref.~\cite{Dokshitzer:2005bf}%
$\,$), will be obtained via next-to-leading order calculations in the 
large-$\nf$ expansion, generalizing the leading-$\nf$ result of Ref.~\cite
{Gracey:1994nn}.
The corresponding contribution to Eq.~(\ref{eq:dP3DMS}) is identical for all
three non-singlet cases and reads
\bea
  \lefteqn{ \delta\, P^{\,(3),\,\rm ns}(x)\,\big|_{\:\nfs} 
  \;\; \equiv \;\; 
           P^{\,(3),\,\rm ns}_{\sigma=1}(x)
         - P^{\,(3),\,\rm ns}_{\sigma=-1}(x)\;\; = \;\; }
  \nonumber\\&& \mbox{\hspn}
  16/81 \: \* \colour4colour{\cfs\,\nfs} \: \*  \Big( \:
       \pqq(x) \: \*  \big[
          - ( 159/4 - 120 \* \z2 + 36 \* \z3 ) \* \H(0)
          + ( 23 + 72 \* \z2 ) \* \Hh(0,0)
          - 279 \* \Hhh(0,0,0)
\nn\\&& \mbox{}
          - 216 \* \Hhhh(0,0,0,0)
          - 76 \* ( \Hh(1,0) + \H(2) )
          - 240 \* ( \Hhh(1,0,0) + \Hh(2,0) + \H(3) )
          - 108 \* ( \Hhhh(1,0,0,0) + \Hhh(2,0,0) 
\nn\\[1mm]&& \mbox{}
            + \Hh(3,0) + \H(4) ) 
       \: \big]
       + (1-x) \: \* \big[ \: 
          - ( 260 - 72 \* \z2 ) \* \H(0)
          - 276 \* ( \Hh(1,0) + \H(2) )
          - 144 \* ( \Hhh(1,0,0) + \Hh(2,0)
\nn\\&& \mbox{}
            + \H(3) )
        \: \big]
       + (1+x) \: \* 
          108 \* \Hhhh(0,0,0,0) 
       - (466 - 398 \* x ) \* \Hh(0,0)
       - (90 - 450 \* x ) \* \Hhh(0,0,0)
         \Big)
\:\: . \quad
\label{eq:dPns3}
\eea
The $\as^{\,n}\,\cfs\, n_{\!f}^{\,n-2}$ contributions dominating
$\delta\, P^{\,(n-1),\,\rm ns}$ in the large-$\nf$ limit are given to all 
higher orders $n$ by a straightforward generalization of Eq.~(\ref{eq:dP2DMS}) 
and the first line of Eq.~(\ref{eq:dP3DMS}). 

Returning to the NNLO coefficient-function and splitting-function differences
(\ref{eq:dC1T}) -- (\ref{eq:ddPqq2}), we note that these functions include
harmonic polylogarithms up to the same weights as the corresponding space-like 
results \cite{Moch:2004pa,Vogt:2004mw,Vermaseren:2005qc,Moch:1999eb}, with the
interesting exception that weight-4 functions enter Eq.~(\ref{eq:dPqq2p}) only
for the SU($n_c$) group-factor combination $\,C_A - 2\,C_F\,$ suppressed as 
$1/n_c$ in the limit of a large number of colours $n_{c\,}$. Except for
the longitudinal coefficient function $c_{L,\rm ns\,}$, the differences between
the time-like and space-like quantities are
parametrically suppressed in the large-$x$ limit, since the logarithmically 
enhanced soft-emission contributions to ${\cal R}_{\:l}$ in Eq.~(\ref{eq:F1dec})
are, as they have to be, invariant under the analytic continuation. 
Specifically, the (identical) leading large-$x$ contributions for
$\,c^{\,(2)}_{T,\,\rm ns} - c^{\,(2)}_{1,\,\rm ns}\,$ and 
$\,c^{\,(2)}_{A,\,\rm ns} - c^{\,(2)}_{3,\,\rm ns}\,$
contain plus-distributions only up to $[(1-x)^{-1} \ln(1-x)]_+$ (all 
proportional to $\pi^2$-terms  arising from the analytic continuation of the 
form factor), and those for the splitting-function differences read, 
\beq
  \delta\, P^{\,(2),\,\rm ns}(x\ra 1) \; = \; 
  -\, 4\: A_{\rm q}^{(1)} A_{\rm q}^{(2)} \ln (1-x) 
  \: + \: {\cal O} (1)
  \:\: , 
\eeq
as predicted in Ref.~\cite{Dokshitzer:2005bf}, where $A_{\rm q}^{(n)}$ are the 
coefficients of $\,\as^{\,n}\, [1-x]_+^{-1}\,$ in Eq.~(\ref{eq:Pns}), 
cf.~Ref.~\cite{Moch:2004pa}. The leading small-$x$ terms, on the other hand, 
differ between the space-like and time-like cases.

\begin{figure}[p]
\vspace{-2mm}
\centerline{\epsfig{file=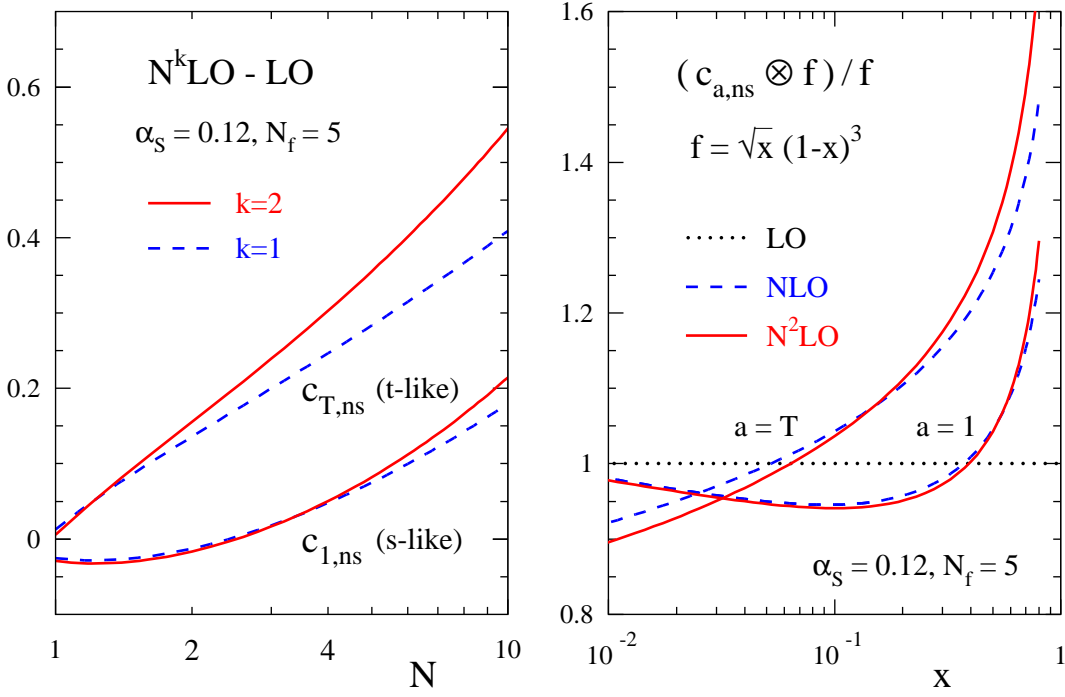,width=14.7cm,angle=0}}
\vspace{-2mm}
\caption{
\label{fig:coeff}
 Comparison of the coefficient function $c^{}_{T,\rm ns}$ for the (time-like)
 process \mbox{$e^+e^- \ra\, h + X$} with its counterpart $c^{}_{1,\rm ns}$ in
 (space-like) deep-inelastic scattering for $Q^2 \simeq M_Z^2\,$.
 Left plot: Mellin moments, right plot: convolutions (\ref{eq:Fah}) with a
 schematic input shape denoted by $f$.}
\vspace{1mm}
\end{figure}

The numerical impact of the second-order contributions to the time-like
coefficient functions $c_a^{(2)}(x)$, $a = T,\:L,\:A$, has been discussed in
some detail already in Refs.~\cite{Rijken:1996vr,Rijken:1996ns,Rijken:1996np}.
Here we confine ourselves to the transverse fragmentation function $F_T$, the
largest contribution to the right-hand-side of Eq.~(\ref{eq:d2sigma}), 
see~Ref.~\cite{Biebel:2001ka}. 
In Fig.~\ref{fig:coeff} we compare the corresponding non-singlet coefficient 
function $c^{}_{T,\rm ns}$ to its counterpart $c^{}_{1,\rm ns}$ in 
deep-inelastic scattering. In order to facilitate a direct comparison, the 
same schematic shape has been used for the non-singlet fragmentation 
distributions and parton distributions. 
As obvious from the figure, the higher-order corrections for $c^{}_T$ are in 
general considerably larger than those for $c^{}_1$. Nevertheless the 
second-order term changes the NLO results, under the conditions of 
Fig.~\ref{fig:coeff}, by 5\% or less from $x \simeq 0.55$ down to very small 
values of $x$.

\begin{figure}[p]
\vspace{-2mm}
\centerline{\epsfig{file=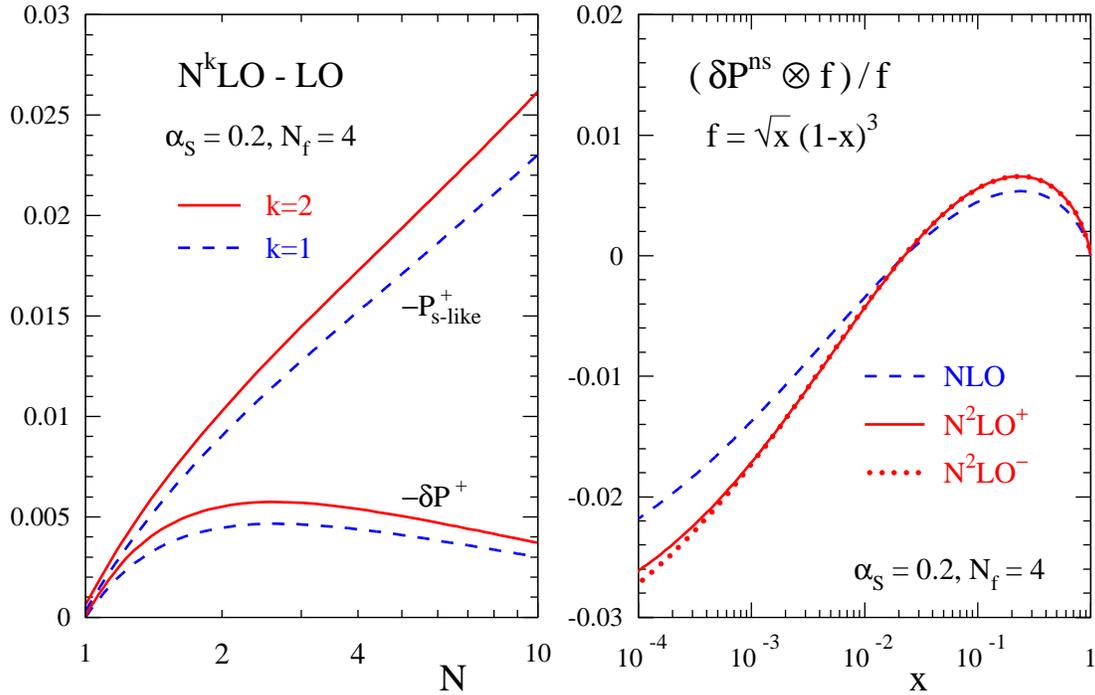,width=14.7cm,angle=0}}
\vspace{-2mm}
\caption{
\label{fig:split}
 The differences $\delta\, P^{\,\rm ns} = P^{\,\rm ns}_{\sigma=1}
 - P^{\,\rm ns}_{\sigma=-1}$ between the time-like ($\sigma=1$) and 
 space-like ($\sigma=-1$) non-singlet splitting functions at a `low' scale
 characterized by the (order-independent) value $\als = 0.2$ of the strong
 coupling constant. Left: moment-space comparison with the higher-order
 corrections in the space-like case. Right: convolutions with a schematic
 input shape.} 
\vspace{1mm}
\end{figure}

The pattern is quite different for the non-singlet splitting functions 
illustrated, in a similar manner but at a lower scale, in Fig.~\ref{fig:split}.
In $N$-space, for example, the ratio 
$\delta P^{\,\rm ns}/(P_{\sigma=-1}^{\,\rm ns} - P^{\,\rm ns}_{\rm LO\,})$
quickly decreases with increasing $N$, at NNLO from about 1/2 at $N=2$ to 
about 1/6 at $N=8$. 
Consequently the total time-like splitting functions $P_{\sigma=1}^{\,\rm ns}$
is only mildly enhanced, e.g., by 8\% and 2\% for these two values of $N$ and
$\als = 0.2$, with respect to their space-like counterparts 
$P_{\sigma=-1}^{\,\rm ns}$ discussed in detail in Ref.~\cite{Moch:2004pa}.
As shown in the right part of Fig.~\ref{fig:split}, the small-$x$ scaling 
violations of the non-singlet fragmentation distributions are weaker than 
those of the parton distributions. For the chosen input distribution this
reduction increases, in a perturbatively stable manner, from about 10\% at 
$x = 10^{-2}$ to about 30\% at $x = 10^{-4}$.

To summarize, we have re-derived the ${\cal O}(\als^2)$ coefficient functions
\cite{Rijken:1996vr,Rijken:1996ns,Rijken:1996np} for the inclusive production
of single hadrons in $e^+e^-$ annihilation \cite{Biebel:2001ka} and obtained,
for the first time, the corresponding third-order splitting functions for the
flavour non-singlet fragmentation distributions. Our derivation of the latter
quantities rests on relations between the time-like and space-like cases, see
especially Refs.~\cite{Curci:1980uw,Floratos:1981hs,Stratmann:1996hn,%
Dokshitzer:2005bf}, and the third-order calculation of deep-inelastic scattering
of Ref.~\cite {Moch:2004pa,Vogt:2004mw,Vermaseren:2005qc}. We expect that a
further study of these relations, backed up by fixed Mellin-$N$ calculations
along the lines of Ref.~\cite{Mitov:2005ps}, will facilitate an extension of 
our derivation to the NNLO flavour-singlet splitting functions and, at least
for $F_L$, the ${\cal O}(\als^3)$ coefficient functions.

Once this step has been taken, the way is open for full NNLO analyses, e.g.,
along the lines of Ref.~\cite{Albino:2005me}, of high-precision data on 
$e^+e^- \!\ra h + X$ from LEP and a future International Linear Collider.
Due to the universality of the splitting functions, our results also represents
a first step towards NNLO analyses of high-$p_T$ hadron production in $ep$
and $pp$ collisions, where very large NLO corrections strongly suggest sizeable
higher-order contributions, see 
Refs.~\cite{Aurenche:2003by,Daleo:2004pn,Kniehl:2004hf}
and \mbox{\cite{Aurenche:1999nz,Kniehl:2000hk,deFlorian:2005yj}}, respectively.
Another application concerns the $b$-quark spectrum in top decays (see 
Ref.~\cite{Corcella:2001hz}$\,$) where, after the calculations of 
Refs.~\cite{Melnikov:2004bm,Mitov:2004du}, the third-order time-like splitting 
functions will facilitate a complete NNLO treatment in the framework of 
perturbative fragmentation.

{\sc Form} and {\sc Fortran} files of our results can be obtained from  
\ {\tt http://arXiv.org} \ by downloading the source of this article.
Furthermore they are available from the authors upon request. 

\vspace*{4mm}
\noindent
{\bf Acknowledgments: }
We are grateful to L. Dixon, P. Uwer and W. Vogelsang for useful discussions.
Our numerical results have been computed using the {\sc Fortran} package of
Ref.~\cite{Gehrmann:2001pz}.
A.M. acknowledges support by the Alexander von Humboldt Foundation.
The work of S.M. has been supported in part by the Helmholtz Gemeinschaft
under contract VH-NG-105.


\end{document}